\documentclass[final]{ustcstep}

\usepackage[dvips]{graphicx}
\usepackage[square]{natbib}
\usepackage{enumerate}
\def\gsim{\;\lower4pt\hbox{${\buildrel\displaystyle >\over\sim}$}\;}
\def\lsim{\;\lower4pt\hbox{${\buildrel\displaystyle <\over\sim}$}\;}
\def\grls{\;\lower4pt\hbox{${\buildrel\displaystyle >\over <}$}\;}

\newcommand\addr[2]{{\footnotesize \it $^{#1}$#2}\\}
\begin{document}

\title{Kinematic Evolution of a Slow CME in Corona Viewed by STEREO-B on 8 October 2007}

\author{Chenglong Shen, Yuming Wang, Bin Gui, Pinzhong Ye, and S.
    Wang\\[1pt]
        \addr{}{KLBPP, School of Earth \& Space Sciences, University of
            Science \& Technology of China, Hefei, Anhui 230026, China;}
    \addr{}{Contact: ymwang@ustc.edu.cn}}

\maketitle
\tableofcontents

\begin{abstract}
We studied the kinematic evolution of the 8 October 2007 CME in the corona based on
\textit{Sun-Earth Connection Coronal and Heliospheric Investigation} (SECCHI) onboard
satellite B of \textit{Solar TErrestrial RElations Observatory} (STEREO). 
The observational results show that this CME obviously deflected to a lower latitude region for about
30$^\circ$ at the beginning.
After this, the CME propagated radially.
We also analyze the influence of the background magnetic field on the deflection of this CME.
We find that the deflection of this CME at an early stage may be caused by the nonuniform distribution of the
background magnetic field energy density and that the CME tended to propagate to
the region with lower magnetic energy density. 
In addition, we found that the velocity profile of this gradual CME shows multiphased evolution during its propagation in COR1-B FOV. 
The CME velocity first kept at a constant of 23.1km.s$^{-1}$. 
Then, it accelerated continuously with a positive acceleration of
$\approx$7.6m.s$^{-2}$.
\end{abstract}

\section{Introduction}
Some parameters of CMEs may greatly change during their propagation in the corona.
The variation of the CME's propagation parameters during this phase
could significantly influence the CME's geoeffectiveness:
(i) the propagation direction variation would determine whether a CME
could arrive on Earth; (ii) the evolution of CME's velocity may change the CME's arrival time.
Thus, the study of the kinematic evolution of the CME in the corona is an important topic
in space physics, especially for space weather study.

The CME deflection in the meridian plane during its propagation in the corona
has been reported by many authors. 
\citet{MacQueen_etal_1986} found that there was an average 2.2$^\circ$ Equator\,--\,ward deflection
of the CMEs in the Skylab epoch (1972 \,--\, 1974) during solar minimum.
\citet{Cremades_Bothmer_2004} studied the differences between the
central position angle (CPA) and the source region position angle
and found that the CME deflected to lower latitude region by about 20$^\circ$ after solar minimum.
This type of deflection would make the CME, which originated from a high latitude region, 
propagate to the ecliptic plane and subsequently encounter Earth.
Based on a STEREO observation, \citet{Kilpua_etal_2009} showed such an example:2 November 2008 CME.
However, the causes for this deflection are still debated.
Previous authors suggested that two factors would cause this deflection of the CME: (i) the influence of
the background coronal magnetic field \citep{MacQueen_etal_1986};
(ii) the fast solar wind flow from polar coronal holes that
encompassed the CMEs' expansion at higher latitude \citep{Cremades_etal_2006}.

Another important factor that would influence the CME's space weather effect
is the velocity evolution of the CME, especially when the CME propagates in the corona 
where the major acceleration of CMEs take place \citep{Zhang_etal_2004}.
Previous results showed that CMEs usually undergo a multiphased
kinematic evolution, including (i) the initiation phase, (ii) the
impulsive (major) acceleration phase and (iii) the propagation
phase \citep{Zhang_etal_2001,Zhang_etal_2004}. However, not all CMEs
necessarily display a full three\,--\,phase evolution. A class
of gradual CMEs characterized by a very weak but long duration
acceleration has also been reported \citep{Sheeley_etal_1999,Srivastava_etal_1999,Srivastava_etal_2000}. 
Based on the different acceleration properties,
\citet{Zhang_etal_2004} 
suggested that there were three classes of CMEs: (i) impulsive acceleration CMEs; (ii) intermediate acceleration
CMEs; (iii) gradual acceleration CMEs.

Recently, the twin \textit{Solar TErrestrial RElations Observatory} (STEREO)
spacecraft \citep{Kaiser_etal_2007}, which was launched on 25 October 2006, 
provide the observations of CMEs with higher resolution and larger range.
 The COR1 instrument \citep{Howard_etal_2008} on SECCHI observed the Sun
from 1.5\,--\,4.0 $R_\odot$ with a time cadence of five minutes, and the COR2 instruments
observed the Sun from 2.0\,--\,15 $R_\odot$ with a time resolution of 15 minutes.
The COR1 design is the first space\,--\,borne internally occulted
refractive coronagraph \citep{Howard_etal_2008} and COR2 imaged the corona 
with five times the spatial resolution and three times the temporal resolution of LASCO/C3
\footnote{\url{http://secchi.nrl.navy.mil/index.php?p=Specifics}}.
Thus, the combination of COR1 and COR2 observations can
be used to study the CME's kinematic evolution in the corona with high spatial and time resolution.

We will study the kinematic evolution including the
propagation direction variation and the velocity evolution of the 8 October 2007 CME 
during its propagation from $\approx$2.0 to $\approx$10$R_\odot$
based on the COR1-B and COR2-B observations (B denotes that it is on the STEREO B spacecraft).
This was a very slow CME with a speed that varied from $\approx$20km.s$^{-1}$ to $\approx$90km.s$^{-1}$ during
its propagation in the COR1-B and COR2-B FOV. 
Because the CME was very slow, COR1-B recorded $\approx$100 images. 
This large number of frames provides us with an opportunity to study the CME's kinematic evolution during its propagation in the corona.
The detailed observations of this CME will be shown in Section 2.
In Section 3 the influence of background magnetic field on the deflection of this CME will be studied in detail by an analytical method.
In the last section, we will give our conclusion and discussion.

\section{Observations}
The CME was firstly recorded by COR1-B at 08:46 UT on 8 October 2007. 
The panels in Figure \ref{def} show the COR1-B and COR2-B observations of this CME. 

Seen from STEREO B, this CME first appeared at solar
west limb. A burst prominence starting at about 07:00 UT on 8 October, 
as seen by EUVI-B 304\AA, was associated with this CME. Because the
CME was launched from the western limb and showed a helical
circular\,--\,like structure, this suggests that the CME was viewed by the
instruments through an axial\,--\,view angle \citep{Cremades_Bothmer_2004,Wang_etal_2009}. Therefore, the projection
of the CME onto the plane of the sky can be treated as the
cross\,--\,section of the CME. \citet{Wang_etal_2009} studied the
internal state of this CME based on its expansions and propagations
from STEREO-B observations.
Note that this slow and weak CME became diffused when it propagated in the later stage of COR1B FOV (from $\approx 3R_\odot$ to $\approx 5 R_\odot$) as shown in Figure \ref{def}(f). Thus, some of the CME's parameters can't be measured correctly
during this period. 
For this reason, these images were excluded from the analysis below.
After the CME propagated into COR2B FOV, 
it represent a clear structure as panels (g)\,--\,(i) in Figure \ref{def} shown. 
Finally, a total of 67 images in COR1 FOV and 14 images in COR2 FOV were used.

\subsection{Deflected Propagation in the Meridian Plane}

\begin{figure*}[tb]
\begin{center}
\includegraphics[width=0.8\hsize]{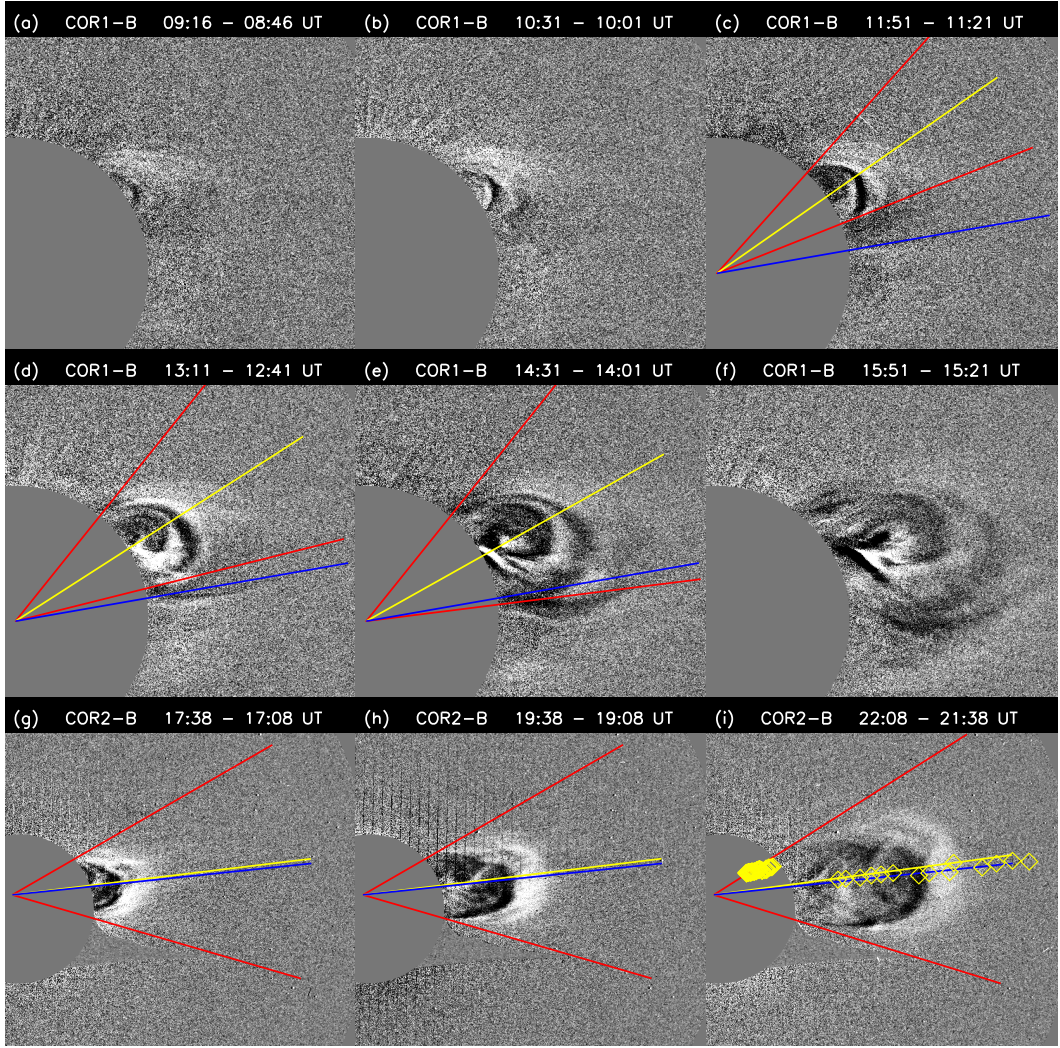}
\caption{Images of 8 October 2007 CME taken by STEREO/SECCHI COR1-B
(a\,--\,f) and STEREO/SECCHI COR2-B (g\,--\,i). The red lines in panel (c)\,--\,(e) 
and (g)\,--\,(i) 
show the selected maximum and minimum position angle of the CME while the yellow lines  
show the radial direction of the CME's CPA. 
The blue lines in these panels show the line with a position angle of 276$^\circ$.
The yellow symbols in panel (i) show the variation of the CPA of the CME.} \label{def}
\end{center}
\end{figure*}

The COR1-B and COR2-B observations of this CME are shown in Panels (a)\,--\,(f) and (g)\,--\,(i) in Figure \ref{def}, respectively.
The red lines in panel (c)\,--\,(e) and (g)\,--\,(i)
show the selected maximum and minimum position angle of the CME while the yellow lines in these panels
show the radial direction of the CME's CPA.
The yellow symbols in panel (i) show the variation of the CPA of the CME. 
The CME is too weak at the beginning as shown in panel (a) and (b) while the CME was diffused at later stage of COR1-B FOV shown as  panel (f) shown. 
So, no position angle was measured during these period.
The panels (a)\,--\,(f) in Figure \ref{def} show that the CME obviously propagated non-radially 
during its propagation in the COR1-B FOV. 
At this stage, this CME deflected to Equator. After the CME propagated into the COR2-B FOV, no obvious
deflection could be found from panels (g)\,--\,(i) in Figure \ref{def}.

\begin{figure*}[tb]
\begin{center}
\includegraphics[width=0.8\hsize]{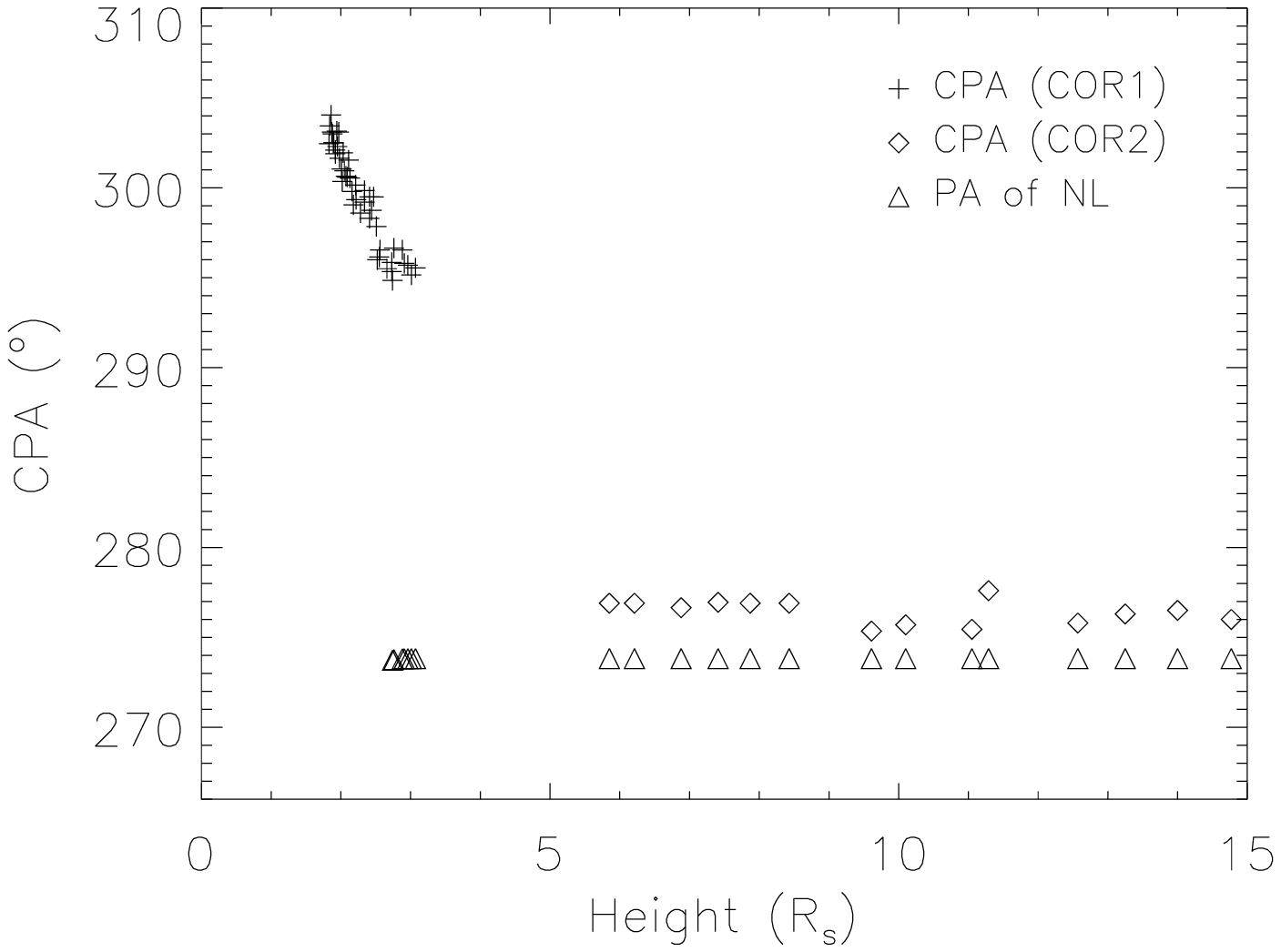}
\caption{CME central angle position (CPA) varied with CME leading
edge height. The crosses and diamonds denote the observations of
COR1-B and COR2-B respectively. The triangles show the position angles
of the heliospheric neutral line at different heights at the same
Carrington longitude as the CME propagation on the meridian plane.}
\label{cpa}
\end{center}
\end{figure*}

Figure \ref{cpa} shows the central position angles (CPAs) of this CME
varied with its leading edge heights (LEHs).
The central position angle (CPA) is defined as the middle position angle
with respect to the two edges of the CME in the sky
plane, and the position angle (PA) is measured counterclockwise from solar North in
degrees \citep{Yashiro_etal_2004}.
The crosses in Figure \ref{cpa} show the COR1-B
observations, while the diamonds show the observations of COR2-B. 
Note that there is a data gap. This data gap appeared because
this slow and weak CME became diffused when it propagated from $\approx$3.0 to $\approx$5.0 $R_\odot$, as we discussed above.
The COR1 observations show that this CME continuously deflected to the Equator at the early stage.
During this stage, the CPA of this CME varied from $\approx306^\circ$ to $\approx294^\circ$.
Then, this CME could continuously deflect to the Equator.
After propagating to  beyond $\approx$5.5$R_\odot$, 
the CME propagated almost radially along the line with a position
angle of $\approx$276$^\circ$ as the blue lines shown in Figure \ref{def}. 
Thus, the observational results show that this CME deflected to the Equator 
at the early stage and then propagated almost radially.

\subsection{Multiple\,--\,phase Evolution of Velocity}
\begin{figure*}[tb]
\begin{center}
\includegraphics[width=0.5\hsize,angle=90]{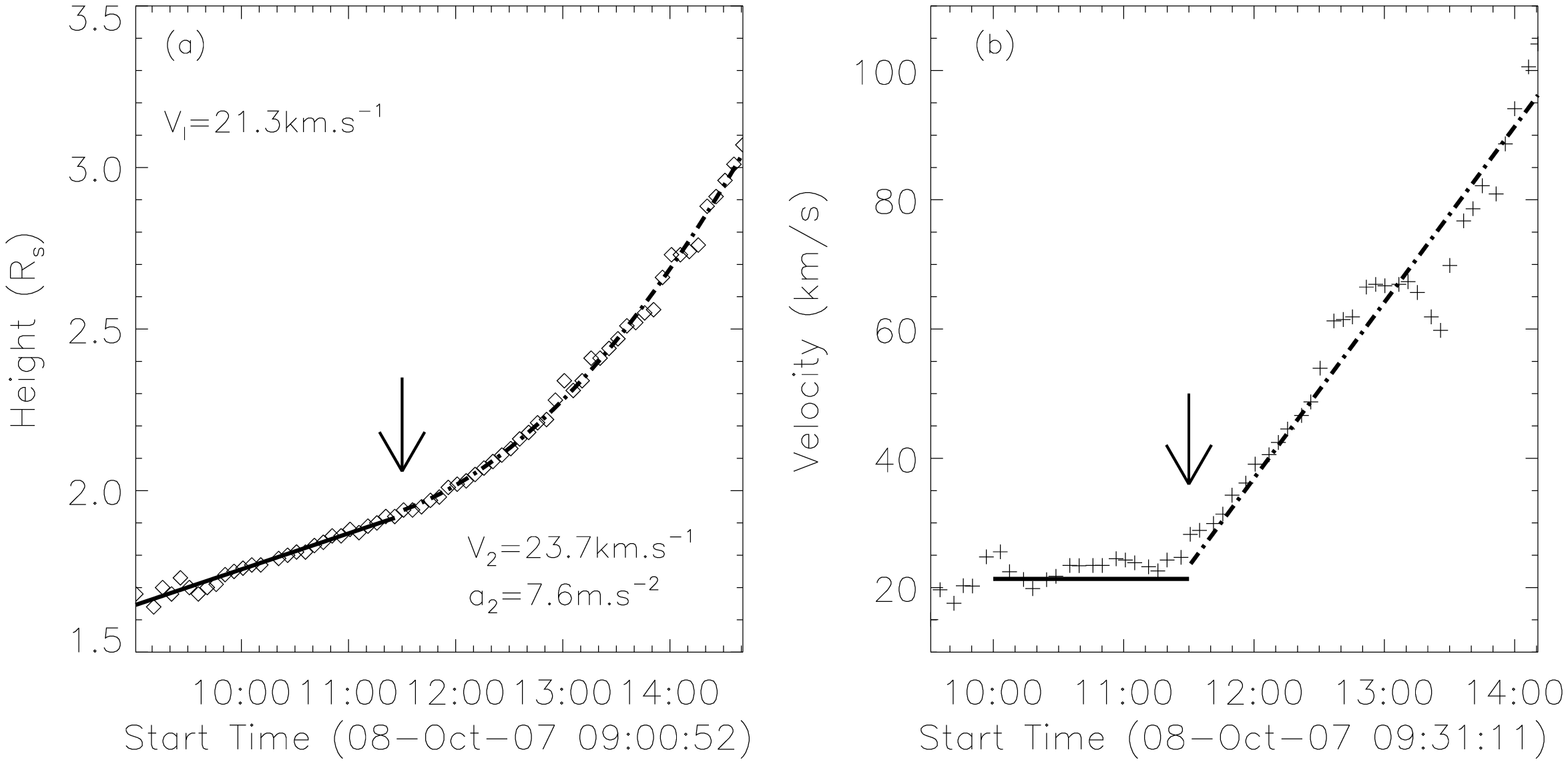}
\caption{Kinematic plots of the 8 October 2007 CME. The left panel
shows the height\,--\,time plot and the right panel show the
velocity\,--\,time plot. The solid lines show the line\,--\,fitting result of
the height\,--\,time distribution of phase 1. The dotted\,--\,dashed lines
show the second order fitting result with constant acceleration in
phase 2. $V_1$ denotes the linear fit velocity of this CME at
the first stage. The $V_2$ and $a_2$ denote the second order fit results
of the velocity and acceleration in second phase.} \label{acc}
\end{center}
\end{figure*}

 \citet{Wang_etal_2009} showed the velocity evolution
of this CME in interplanetary space from its burst to the propagation to
the place beyond 70 $R_\odot$. This is a gradual CME with a weak but
long\,--\,duration acceleration. The list of the solar event
report\footnote{\url{http://www.swpc.noaa.gov/ftpdir/warehouse/}}
generated by NOAA/SEC indicated that no major solar flare was associated
with this event. It was consistent with the result suggested by
\citet{Zhang_Dere_2006} that CMEs without flares usually show
gradual properties with a weak but long\,--\,duration acceleration.

Based on the COR1-B observations, the height\,--\,time and
velocity\,--\,time evolutions of this CME during its propagation in the
COR1 FOV are shown by diamonds in Figure \ref{acc}(a) and crosses
in Figure \ref{acc}(b) respectively. The velocity measurements in
Panel (b) were obtained by using the twelve\,--\,points running
straight\,--\,line fitting method on HT measurements similar as in
\citet{Sheeley_etal_1997}. At the beginning, the velocity of
this CME is kept at a constant before the time of $\approx$11:30 UT (phase 1).
The hight\,--\,time measurements during this phase can
be well fitted by a straight line.
The fitting result of the velocity at this phase is
$\approx$23.1km.s$^{-1}$.
After the time of 11:30 UT, the speed of the CME continuously and gradually increased (phase 2).
The dotted\,--\,dashed lines in Figure \ref{acc} show
the second order fitting result based on the height\,--\,time measurements with a
constant acceleration. The fitting result correlates well with the observations.
The acceleration of the CME during this
phase was 7.6m.s$^{-2}$. This velocity profile shows that a CME
with this gradual acceleration goes through a multiphased
velocity evolution.

\section{The Influence of the Background Magnetic Field on the CME's Deflection}
\subsection{General Idea}
The deflected propagations of CMEs and their driven shocks could be interpreted as the results of
the interaction of the CMEs with the background solar wind and the magnetic field
\citep[e. g.][]{MacQueen_etal_1986,Wei_Dryer_1991,Gopalswamy_etal_2004}).
So far, no theoretical analysis has been done to quantitatively study 
how the direction and magnitude of these deflections are influenced by background magnetic field. 
Therefore, we propose a simple method to discuss the CME's deflection. 
Consider that a CME propagates in the coronal medium,
the background magnetic field is perturbed by the CME,
and the field lines that originally were in the region occupied by the CME are expelled.
Compared with the unperturbed corona, the expelled magnetic field lines go around the CME and become significantly compressed,
which means lots of free magnetic energy is built-up.
The free energy from the compression provides a restoring force that acts on the CME.
Since the background magnetic field is not uniformly distributed,
the restoring forces acting on the different areas of the CME should not be the same.
Therefore, the deflection may happen through the effect of the resultant force. 

\begin{figure*}[tb]
\begin{center}
\includegraphics[width=0.6\hsize]{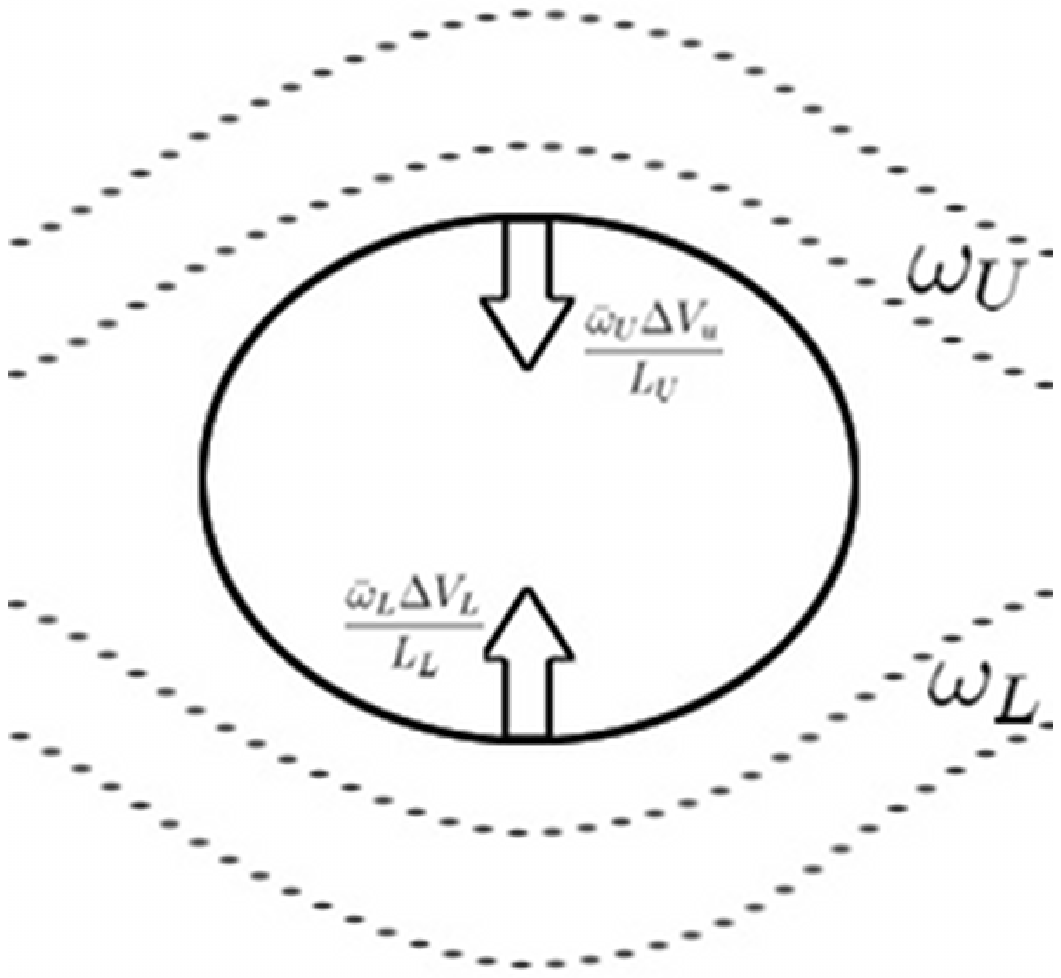}
\caption{Sketch map of the CME\,--\,perturbed background magnetic
field.} \label{sketch}
\end{center}
\end{figure*}

Consider a simple case as shown in Figure \ref{sketch}, in which we study the region occupied by the CME
as an upper part and a lower part.
The free energy corresponding to the upper part can be approximated as the original magnetic energy in it,
\textit{i.e.}, $W_U\approx\int \omega \mathrm{d}x^3\approx \overline\omega_U \Delta V_U$,
where $\omega=\frac{B^2}{2\mu}$ is the energy density and $\overline\omega_U$ is the average energy density
in the upper part. The associated restoring force acting on the upper part points downward and is roughly given
by $f_U \approx \frac{W_U}{L_U}$, where $L_U$ is the characteristic length of the part.
Similarly, there is a restoring force acting on the lower part, which points in the opposite direction.
Comparing the two forces, we can estimate the direction toward which the CME will propagate.

The above scenario can be easily extended to a more general situation.
The net force acting on the CME is $f=f_U-f_L\approx\frac{W_U}{L_U}-\frac{W_L}{L_L}$. Obviously, its generalized form is
\begin{eqnarray}
\textbf{f}=-\nabla\omega=-\nabla\left(\frac{B^2}{2\mu}\right). \label{eq_deflection}
\end{eqnarray}
Since we only know the CME's deflected propagation in the latitudinal direction, it is not necessary to use Equation (\ref{eq_deflection}), 
and we just apply it as in the simple case given in the last paragraph.

Furthermore, we simply assume $\Delta V_U=\Delta V_L=\Delta V$ and $L_u=L_L=L$.
Accordingly, the net force that acts on the CME can be rewritten as
\begin{eqnarray}
f=\frac{\Delta V}{L}(\omega_U-\omega_L)=\frac{\Delta V}{L}\Delta\omega,\label{eq_2}
\end{eqnarray}
where $\Delta \omega=\omega_U-\omega_L$ is the difference of the magnetic energy density between 
the upper and lower part.
Based on Equation (\ref{eq_2}),
a positive value of $\Delta \omega$ indicates that the CME\,--\,perturbed magnetic
field will act an Equator\,--\,ward force on the CME, while a negative value
of $\Delta \omega$ means an polar\,--\,ward force would act on the CME.

\subsection{Calculated \textit{vs.} observed Results}

\begin{figure*}[tb]
\begin{center}
\includegraphics[width=1\hsize]{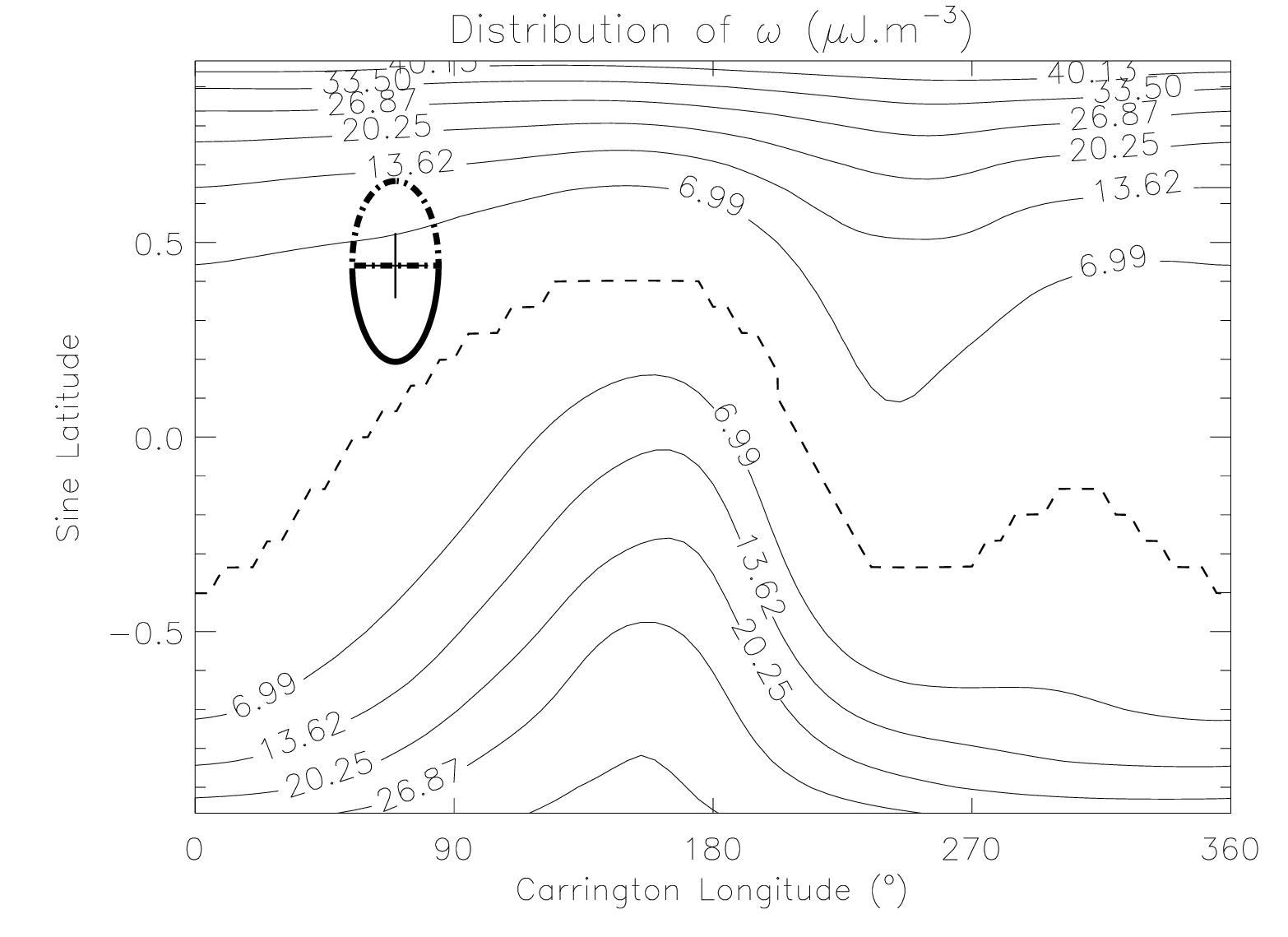}
\caption{Distribution of the magnetic energy density ($\omega$)
of the Carrington rotation 2061 at the hight of 3 $R_\odot$. The
dashed line shows the neutral line at this height. The black cross
shows the position of this CME at the hight of 3 $R_\odot$. The regions
enclosed by the dotted\,--\,dashed line and the solid line shows the selected upper and lower part, respectively.} \label{pdis}
\end{center}
\end{figure*}

To apply this method, the background magnetic energy density should be given.
However, there is no direct measurement of the solar magnetic field except at the photosphere up till now.
The three\,--\,dimension(3D) magnetic field of the solar corona could only be calculated by different models.
In this work, the magnetic energy density will be calculated from the 3D magnetic field extrapolated by
the Current Sheet\,--\,Source Surface (CSSS) model. 
The CSSS model was developed by Zhao and his colleagues \citep{Zhao_Hoeksema_1995,Zhao_etal_2002a}.
This model has been used to reproduce the magnetic field strength at 1AU \citep{Zhao_etal_2002a} and calculate
the strength of CME\,--\,driven shock in the corona \citep{Shen_etal_2007}.
We adopt the bottom boundary from the WSO (Wilcox Solar Observatory)
synoptic charts\footnote{\url{http://wso.stanford.edu/synopticl.html}},
which are assembled from individual magnetograms observed over a Carrington rotation.

Figure \ref{pdis} shows the magnetic energy density distribution of
the Carrington rotation 2061 at the height of 3 $R_\odot$. The
Carrington rotation 2061 corresponds to the period from 20:21 UT 10
September 2007 to 02:58 UT 8 October 2007. We use the previous
synoptic chart instead the one covering the CME onset time,
because the time of the CME propagation meridian passing across the
central meridian was in Carrington rotation 2061.
From Figure \ref{pdis} an obvious nonuniform distribution
of the magnetic energy density can be found.
The dashed line in Figure \ref{pdis} shows the neutral line.
The neutral line is defined as the boundary between the negative and positive radial magnetic field and 
corresponds to the position of the Heliospheric Current Sheet(HCS) at that height. 
As in previous results, the magnetic field increased gradually above and below the HCS \citep{Wolfson_1985} and
the radial field was generally equal on the two sides \citep{Burton_etal_1996} as shown in Figure \ref{pdis}.
From Figure \ref{pdis} it is obvious that the direction of the magnetic energy density gradient is toward to the HCS, where the
magnetic energy density is lowest. 

The triangles in Figure \ref{cpa} show the position angles of the neutral lines 
at the CME propagation meridian for different heights. 
It is obvious that this CME deflected to the HCS at the beginning and then propagated almost along it.
As the HCS located at near the solar Equator region during this period, 
this CME showed an obvious Equator\,--\,ward deflection.

\begin{figure*}[tb]
\begin{center}
\includegraphics[width=0.8\hsize]{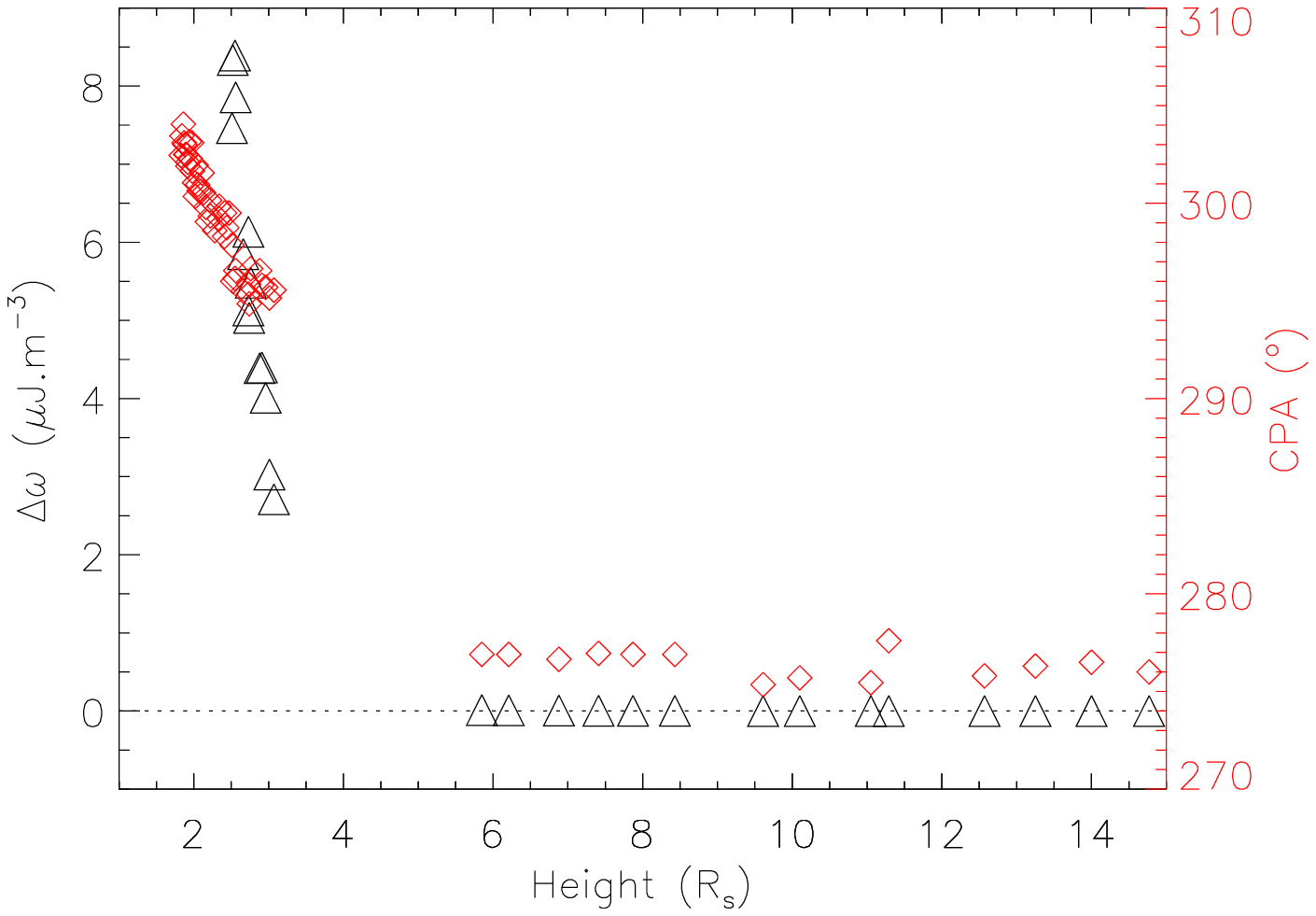}
\caption{$\Delta \omega$ varied with leading edge height. The
black triangles show the calculated result of $\Delta \omega$ at different altitudes.
    The red symbols show the CPA varied with the LEH of the CME as in Figure \ref{cpa}.} \label{deltap}
\end{center}
\end{figure*}

We applied  Equation (\ref{eq_deflection}) to analyze the
 variation of $\Delta \omega$ during the 8 October 2007 CME's propagation in the corona.
The regions enclosed by the dotted\,--\,dashed line and the solid line in Figure \ref{pdis} 
show the selected upper part and lower part in a 15$^\circ$ angle range. 
The enclosed circle covers a 30$^\circ$ angle range and corresponds to the mean angle width of this CME
during its propagation in COR1-B FOV.
Figure \ref{deltap} shows the variation of $\Delta\omega$ and CPA with the CME's leading edge heights.
Shown as black triangles in Figure \ref{deltap}, the $\Delta \omega$ were
positive for almost all heights. 
As we discussed above, this result means that an Equator\,--\,ward force was acting on the 8 October 2007 CME.
This Equator\,--\,ward force could deflect the CME to a lower latitude. 
It is well consistent with the CPA observations shown as the red diamonds in Figure \ref{deltap}. 
During the CME's outward propagation and its Equator\,--\,ward deflection, 
the value of $\Delta \omega$ decreased quickly.
When the CME propagated to $\approx5.5R_\odot$, $\Delta \omega$ almost reached the value of $\approx$0 and then stayed at $\approx$0. 
If $\Delta \omega$ equals zero, this means no obvious resultant force
would act on the CME. 
Accordingly, it is quite natural that the CME propagated radially during this period, as the observations show.
These results illustrate that the deflection of the CME was controlled by the background magnetic field 
and the gradient of magnetic energy density, which made the CME deflect to the region 
with lower magnetic energy density.

\section{Summary and Discussion}
We studied the kinematic evolution of the 8 October 2007 CME during its propagation in the corona. 
From the study of the CME's propagation direction evolution in the meridian plane, 
we found that this CME obviously deflected to Equator at an early stage and then propagated almost
radially. Combined with the three\,--\,dimension coronal magnetic field extrapolated from the CSSS model,
we found that the CME deflected to the HCS at the beginning and then propagated almost along the HCS. 

The velocity evolution of the CME during its propagation 
in the COR1-B FOV was also studied. 
The observations show that this CME underwent a two\,--\,phase evolution:
\begin{enumerate}[(i)]
\item At the beginning, the CME propagated outward with a constant speed of $\approx$23. km.s$^{-1}$. 
This constant speed phase caused a velocity plateau in the velocity profile. 
Such a velocity plateau has also been reported and discussed by other authors
\citep{Dere_etal_1999,Srivastava_etal_1999,Zhang_Dere_2006}.
\citet{Dere_etal_1999} suggested that it might correspond to the
period when the CME was opening the helmet streamer field lines.
\item After the velocity plateau, the speed of the 8 October 2007 CME continuously increased with an acceleration of 7.6$m.s^{-1}$.
\end{enumerate}
These results show that this gradual CME experienced a multiphased
acceleration.  \citet{Srivastava_etal_1999} also found that in some
cases of a gradual CME that they studied, a two\,--\,step speed
profile occurred. They found that the CME accelerated initially
until reaching a certain height and then moved with an almost
constant speed for some time, before finally drifting away with the
slow solar wind. 

Furthermore, we performed a theoretical analysis to quantitatively study the CME's deflection.
We found that the deflection of the CME was controlled by the background magnetic field and 
that the CME tends to deflect to the region with lower magnetic energy density.
We recall that we only consider the distribution of the magnetic
energy at a given height, while the difference along the radial direction
is neglected. 

\begin{figure*}[tb]
\begin{center}
\includegraphics[width=1\hsize]{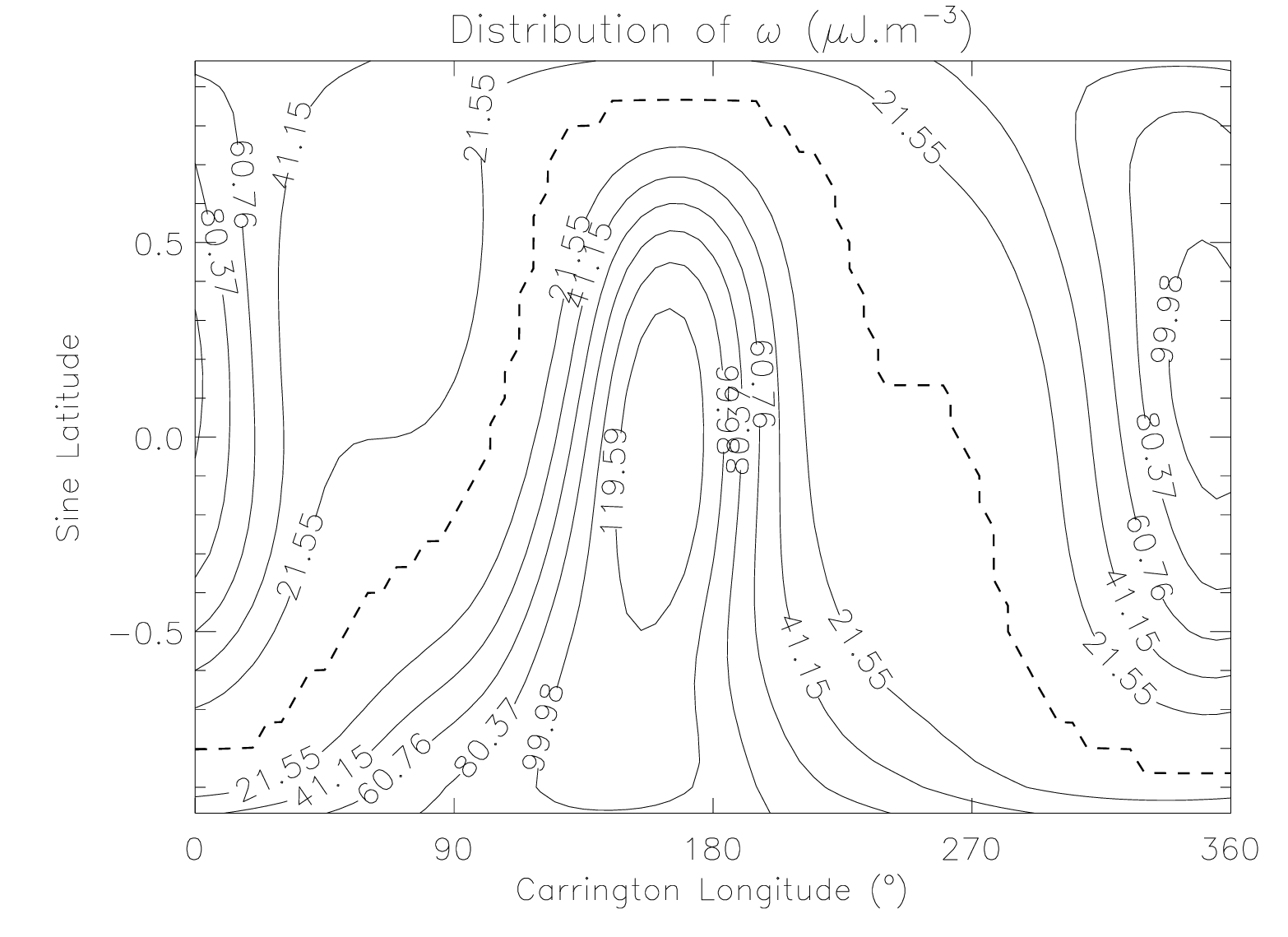}
\caption{Distribution of the magnetic energy density of the
Carrington rotation 1978 at the height of 3 $R_\odot$. The dashed
line shows the neutral line.} \label{pdis1978}
\end{center}
\end{figure*}

In Section 3 we analyze the influence of the background magnetic field on the CME's deflection
from the point of view of the energy. From a different point of view,
the electromagnetic force of the perturbed magnetic field acting on the CME consists of two components: 
the magnetic pressure force $f_p=\frac{\nabla B^2}{2\mu}$ and the magnetic
tension force $f_t=\frac{\nabla\cdot B B}{\mu}$. 
The direction of the magnetic pressure force is from the higher magnetic field strength region to 
the lower magnetic field strength region.
Besides, when a CME propagats outward, it  compresses the background magnetic field lines as shown
in Figure \ref{sketch}. In this situation, the magnetic tension force at a given point may be contrary to the
compression direction. Because the magnitude of the tension force is related with the value of
magnetic field strength, the resulting tension force that acts on the CME may also be directed toward the lower
magnetic field strength region. Combined with the effect of magnetic pressure force and magnetic tension force,
one can expect that a force directed toward the lower magnetic field strength region would act on the CME when it propagated outward
and the CME tends to deflect to the lower magnetic energy density region.

Previous statistical results showed that the CME did deflect to the Equator at solar minimum, 
but no obvious regularity at solar maximum could be found \citep{MacQueen_etal_1986,Cremades_Bothmer_2004,Cremades_etal_2006}.
 In the solar minimum the current sheet that corresponds to the lower magnetic energy density region
is usually located near the Equator \citep[e. g.][]{Shodhan_etal_1994} similar to that we shown in Figure \ref{pdis}.
Based on the result that the CME would deflected to the lower magnetic energy density region, 
the previous statistical results that the CME deflected to the Equator at solar minimum can be explained.
 The CME studied here happened on 8 October 2007,  during the deep solar minimum of the 23rd cycle. 
It is a good example of the CME's Equator\,--\,ward deflection in the solar minimum influenced by the background magnetic field. 
In the solar maximum, the distribution of the magnetic energy density was not as simple. 
Figure \ref{pdis1978} shows an example of the magnetic energy density distribution
at the height of 3 $R_\odot$ of the Carrington rotation 1978,
which corresponds to the period from 03:13 UT 30 June 2001 to 08:08 UT 27
July 2001, during the solar maximum of the 23rd solar cycle. We found
that the magnetic energy density distribution at solar maximum was
complex and that no definite low magnetic energy density region where the CME would deflect to could be expected.
Therefor, no obvious regularity of the deflection of a CME during solar maximum could be explained as the previous results showed.

The nonuniform distribution of the magnetic energy density may also exist in the longitude direction, as shown in Figure \ref{pdis} and Figure
\ref{pdis1978}.
This nonuniform distribution was more obvious at solar maximum (shown in Figure \ref{pdis1978}). 
Thus, it could be expected that the CME would deflect to the east or west of the ecliptic plane because of the
gradient of the magnetic energy density especially in solar maximum.
This may provide another mechanism that could trigger the CME's East\,--\,wWest
deflection besides the interplanetary spiral magnetic field
\citep{Wang_etal_2004b,Wang_etal_2006} and coronal holes \citep{Gopalswamy_etal_2004,Gopalswamy_etal_2005b,Gopalswamy_etal_2009}.
The East\,--\,West deflections of the CME in ecliptic plane
can influence the CME's geoeffectiveness \citep{Wang_etal_2006,Lugaz_etal_2010}. 
It was thought to be the cause of the East\,--\,West asymmetry distribution of earth\,--\,encountered CME's source region \citep{Wang_etal_2002b,Zhang_etal_2003}.


Previous studies also suggested that the heliospheric magnetic field
(HMF) would influence the CME and the shock's
propagation \citep{Wei_Dryer_1991,Smith_2001,Yurchyshyn_etal_2007,Yurchyshyn_2008,Xie_etal_2006}.
By analyzing flare\,--\,associated shock wave events based on
interplanetary scintillation (IPS) observation,
\citet{Wei_Dryer_1991} found that all flare\,--\,associated shock
waves tended to propagate toward the low latitude region near the
solar Equator and the propagation directions tended toward the HCS.
They suggested that this was caused by the dynamic action of near\,--\,Sun
magnetic force on the ejected coronal plasma.
\citet{Yurchyshyn_2008} found that the tilt of the coronal neutral
line axis corresponded well with the magnetic cloud (MC) axis
orientation observed at 1AU, but found no agreement with the EIT arcades.
He suggested that the ejecta might be rotated in a way that it
locally aligns itself with the heliopsheric current sheet. The
HMF may have deflected the CME from its initial direction and, quite
possible, rotated the axis of the CME as it moved through
interplanetary space \citep{Smith_2001}. We showed here the whole
process of the CME deflection to the HCS influenced by the heliospheric
magnetic field and analyzed it for a very slow CME
event. These results indicate that the heliospheric magnetic field
could significantly influence the CME and its connected events.

We acknowledge the use of the data from STEREO/SECCHI.
We thank Jie Zhang for helpful discussions.
 This work is supported by grants from the National
Natural Science Foundation of China (40904046,40874075,40525014), the 973
National Basic Research Program (2011CB811403), Ministry of
Education of China (200530), the Program for NewCentury Excellent
Talents in University (NCET-08-0524), the Chinese Academy of
Sciences (KZCX2-YW-QN511, KJCX2-YW-N28 and the startup fund)
and the Fundamental Research Funds for the Central Universities.


\end{document}